# Naïve Physics and Quantum Mechanics: The Cognitive Bias of Everett's Many-Worlds Interpretation

Andrew SID Lang* and Caleb J Lutz

*Department of Computing and Mathematics, Oral Roberts University, USA*

*Address correspondence to this author at the Department of Computing and Mathematics, 7777 S. Lewis Ave., Tulsa, OK, 74171 USA. E-mail: alang@oru.edu

**Keywords:** cognitive bias, Everett's interpretation, naïve physics, quantum mechanics

**Abstract:** We discuss the role that intuitive theories of physics play in the interpretation of quantum mechanics. We compare and contrast naïve physics with quantum mechanics and argue that quantum mechanics is not just hard to understand but that it is difficult to believe, often appearing magical in nature. Quantum mechanics is often discussed in the context of "quantum weirdness" and quantum entanglement is known as "spooky action at a distance." This spookiness is more than just because quantum mechanics doesn't match everyday experience; it ruffles the feathers of our naïve physics cognitive module. In Everett's many-worlds interpretation of quantum mechanics, we preserve a form of deterministic thinking that can alleviate some of the perceived weirdness inherent in other interpretations of quantum mechanics, at the cost of having the universe split into parallel worlds at every quantum measurement. By examining the role cognitive modules play in interpreting quantum mechanics, we conclude that the many-worlds interpretation of quantum mechanics involves a cognitive bias not seen in the Copenhagen interpretation.

## Introduction

Neils Bohr said of quantum mechanics, "Those who are not shocked when they first come across quantum theory cannot possibly have understood it." [1]  When one first learns about quantum mechanics, it appears to be a theory that cannot be describing reality. In fact, the more one learns about it, the more counterintuitive it becomes, seeming supernatural. While the theory is mathematically elegant and the predictions of the theory have been tested and verified, there is something about quantum mechanics that just doesn't feel right. Can the underlying reality of nature really be so bizarre?

Presently, the two prevailing interpretations of quantum mechanics are the *Copenhagen interpretation*, developed by Bohr and Heisenberg in Copenhagen in the 1920s [2], and the *many-worlds interpretation*, first proposed by Hugh Everett in 1957 [3].  Whereas the Copenhagen interpretation explains the surprising outcomes of quantum theory by positing that nature is inherently non-local, the many-worlds interpretation tries to recover from this counter-intuitiveness (and the need for a mechanism for wave-function collapse) at the expense of postulating that the universe is constantly splitting apart at every quantum interaction in the universe.

This paper proposes that the adherence to the many-worlds interpretation is biased by the human perception of reality arising from an innate understanding of physics. That is, the Copenhagen interpretation's non-locality feels wrong and at times mystical, making the many-worlds interpretation more appealing because of our innate cognitive structure.

## Cognitive Modules

Since at least the time of Plato and Aristotle, the debate over whether humans are inborn with inherent knowledge and even a knowledge-processing system has fascinated the world's greatest thinkers. While at first these types of arguments were limited primarily to philosophers, modern research has spurred this discussion into the systems of education and science, particularly now preoccupying the work of many scholars in the field of cognitive psychology. While for centuries, philosophers and early psychologists argued between potential cognitive inheritances from *Tabula Rasa* to *innatism*, only in the last half century has empirical data arisen validating the existence of an inborn psyche.

These predisposed thought structures found common in newborns and very young children, referred to as a child's *cognitive modules*, give them an innate knowledge that allows them to process new information in certain ways [4]. Research has shown that cognitive modules appear to be involved in a child's ability to, amongst other things, process and ascertain language [5], distinguish the self from others [6], and rely on object permanence [7]. Early research by Piaget [8] demonstrated how cognitive modules lead people to systematically view the world in biased ways, an effect called *cognitive bias*. Cognitive biases may not match reality. For example, these cognitive biases can cause people to systematically make errors when interpreting theories of motion [9]. Thus, it seems reasonable to assert that naïve physics, defined by Proffitt as "the commonsense beliefs that people hold about the way the world works" [10], may have similarly biased the interpretation of quantum mechanics.

## Naïve Physics

Research has shown that many core beliefs are solidified in children either at a very young age or else are entirely inborn. When analyzing intuitive beliefs about motion it has been found that they have more in common with medieval (Aristotelian) mechanics rather than classical Newtonian mechanics.

Nersessian and Resnick present the intuitive beliefs about motion as follows [11]:
1. All motion requires a causal explanation.
2. Motion is caused by a "force."
3. Continuing motion is sustained by a stored "force."
4. Active and passive motions differ.
5. Downward motion is natural.
6. Heavier objects fall faster.

A child's inherent belief in a stored force or *impetus* has been observed in multiple studies, sometimes even leading to false conclusions. For example, the *curvilinear bias* is a misconception where objects constrained to a curved path are assumed to continue to follow a curved path once released [12].  Contrary to our innate curvilinear bias, objects released from a curved path move in a line tangent to the release point rather than continue to follow a curved path.

Fascinatingly, many facets of naïve physics such as the curvilinear bias are often seen pervading into adulthood.  A curvilinear bias is seen in the misrepresentation of physical properties in the thriller *Wanted* where the main characters are capable of sending bullets on curved paths simply by firing while swinging their arm [13]. Though tantalizing one's inborn cognitive disposition, this action defies physics.  This Hollywood movie is just one illustration of how childhood judgements of physical properties can carry on into adulthood.  Perhaps even more fascinating is the fact that even those "trained in physics" often revert back to evaluating via their cognitive biases if required to answer physics riddles under time pressure [10, 14, 15].

Another famous example of children's cognitive biases in physics that can mislead has become known as the *gravity error*; the error arising from the innate belief that "downward motion is natural."  Explored by psychologist Bruce Hood, this error comes from children not understanding that certain laws can modify the law of gravity.  Hood found that, when viewing a ball dropped into one of three holes, young children always look in a pocket directly beneath the original hole in which the ball was originally dropped.  These children did not have to be trained to do this: they appeared to instinctively know that objects fall straight down.  An interesting twist on the experiment showed that children continue to do so even when a tube is connected to the mouth of one hole and stretched to a pocket other than the one directly below.  The infants will still look directly beneath the hole in which the ball was dropped and ignore completely the pocket to which the tube stretches.  This makes it seem that infants do not understand that a curving tube can override the law of gravity [16].

Other examples of these naïve understandings of physics are abundant.  When a tennis ball is thrown into the air, we expect it to come back down.  When one object rolls into another, we expect it to stop or bounce off, not pass right through.  Bruce Hood, in his book *SuperSense* [17], expounds upon four rules of naïve physics that underlie all physical intuitive reasoning, proposed by infant psychologist Elizabeth Spelke [18].  These four rules are pivotal in our discussion of the effect cognitive biases have on theories of quantum mechanics, and as such, will be quoted below as presented by Hood in his book [17]:

Rule 1: *Objects do not go in and out of existence like the Cheshire Cat in Alice in Wonderland. Their solidity dictates that they are not phantoms that can move through walls.  Likewise, other solid objects cannot move through them.*
Rule 2: *Objects are bounded so that they do not break up and then come back together again. This rule helps to distinguish between solid objects and gloop such as applesauce or liquids.*
Rule 3: *Objects move on continuous paths so that they cannot teleport from one part of the room to another part without being seen crossing in between.*
Rule 4: *Objects generally only move when something else makes them move by force or collision.  Otherwise, they are likely to be living things…*

As we shall see later, Spelke's rules are remarkable to someone familiar with quantum mechanics because they are almost exactly the opposite of the phenomena we observe in the quantum realm.

Many cognitive biases, including those detailed by Spelke, persist into adulthood, yet most naïve physical beliefs are also modified over time: an adult viewing Hood's gravity error experiments would most likely pick the pocket located at the end of the tube, not directly under the original hole.  It seems that humans can consciously overcome our inbuilt laws of physics, but much evidence shows that at the core there still remains the four fundamental beliefs espoused by Spelke [18].  Most grown adults would heartily agree that objects do not go in and out of existence, are not shattered only to reconfigure perfectly, do not have invisible teleportation abilities, and only move when acted upon.  Indeed, appearances violating these beliefs, such as disappearing rabbits and levitating tables, are usually regarded as magical [17].

**Quantum Mechanics is Spooky**
Over the last two-hundred years quantum mechanics has developed into a rich, beautiful and compelling theory. However, the way physical objects behave in the quantum realm is very different both from what we observe in every day experience and from our naïve understanding of physics.  Taking Spelke's rules of innate object knowledge and comparing them to quantum mechanical phenomena, we see stark differences:

Rule 1: *Objects do not go in and out of existence like the Cheshire Cat in Alice in Wonderland. Their solidity dictates that they are not phantoms that can move through walls.  Likewise, other solid objects cannot move through them.*

In quantum mechanics, particles can appear to go in and out of existence! Quantum mechanics predicts that all around us and all the time particles called virtual particles and constantly popping in and out of existence [19].  Interestingly enough, Timothy Ferris independently referred to these unintuitive quantum mechanical results as "Alice-in-Wonderland oddities" [20].

Similarly, particles can move through walls. Classically, when particles collide with a barrier, with not enough energy to burst through, they bounce back – they *reflect*; yet in quantum mechanics, sometimes particles with not enough energy to overcome a barrier still make it through via a process known as *quantum tunneling,* see figure 2.

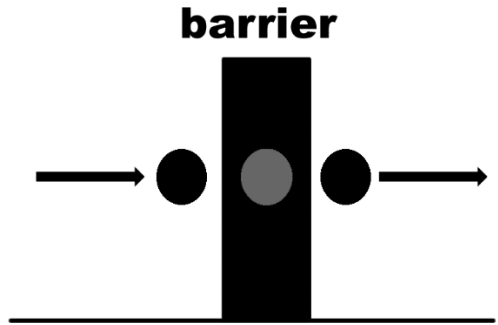

**barrier**

Figure 1. Quantum tunneling through a barrier.

Rule 2: *Objects are bounded so that they do not break up and then come back together again. This rule helps to distinguish between solid objects and gloop such as applesauce or liquids.*

When a single photon hits a half-silvered mirror, see figure 2, intuition would say that the photon randomly leaves the mirror in either the vertical or horizontal direction. The photon would then bounce off one of the mirrors and then hit the half-silvered mirror just before the detectors. At the second half-slivered mirror one would again expect the photon to randomly leave in either the vertical or horizontal direction and thus we should detect half the photons at detector A and half the photons at detector B.

However, when we perform these experiments, we observe 100% of the photons arriving at detector A and none at detector B. This effect is known as *one-particle interference*. Even when only a single photon is sent through the apparatus, it appears as if the photons splits, travels along both paths, and comes back together again to interfere with itself.

If we choose to observe the path the photon travels along, we do indeed only see it traveling along one path, as one would expect, but interestingly the interference effect no longer occurs; we see equal amounts of photons arrive at both detectors. The one-particle interference effect only happens when we're not looking. Strangely, this choice to look or not can be made after the photon has completed its journey through the apparatus with the same effect [21, 22].

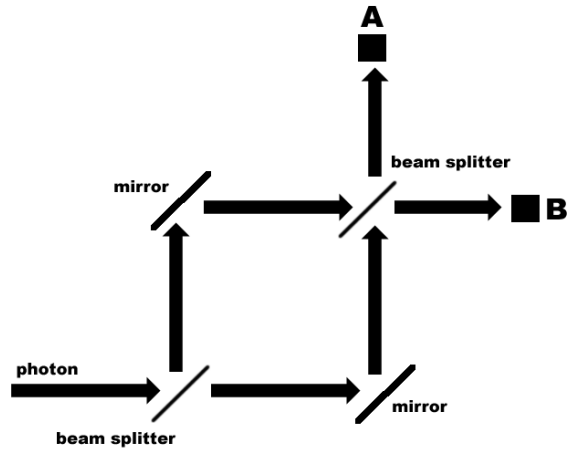

Figure 2. Photon seems to split apart and travel along both paths.

Rule 3: *Objects move on continuous paths so that they cannot teleport from one part of the room to another part without being seen crossing in between.*

In quantum mechanics, particles can teleport from one location to another without crossing in between. For example, if a particle is constrained to be within a box with walls that are impenetrable barriers (so it can't tunnel out), then there are certain states where a particle can be observed on the left hand side of the box and on the right hand side of the box but never in the middle. Sometimes when you look, you see the particle on the left hand side of the box; sometimes when you look, you see the particle on the right hand side of the box; but you will never observe the particle in the middle travelling from one side to the other as our intuition demands, see figure 3.

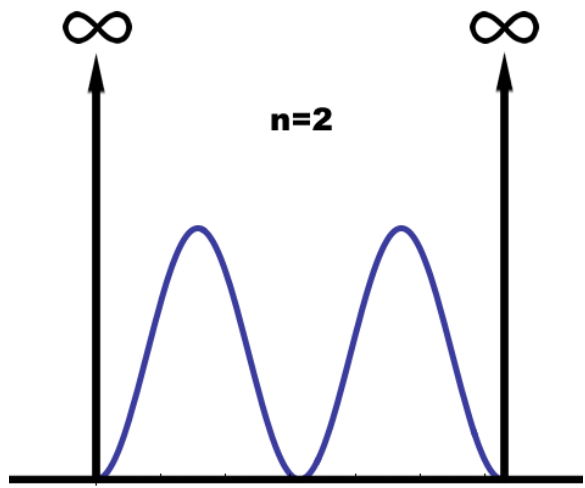

Figure 3. *Particle in a Box.* The probability of finding the particle in the middle is zero.

Rule 4: *Objects generally only move when something else makes them move by force or collision. Otherwise, they are likely to be living things…*

In quantum mechanics, all particles, even those that are in the state of lowest possible energy — *ground state* — still have a *zero-point energy*. This underlying energy causes all particles to be in constant motion without being acted upon by external forces or collisions, even at absolute zero.

As we can see, quantum phenomena do not just go against everyday experience, they go against our inbuilt cognitive modules. This not only makes quantum mechanics difficult to comprehend but also difficult to accept. For those trained in quantum mechanics, it is even worse. The Copenhagen interpretation postulates that not only are the results of quantum mechanics non-local but that reality itself is non-local. Feynman summed this up by saying "… the 'paradox' is only a conflict between reality and your feeling of what reality 'ought to be'." [23]

Other quantum phenomena that seem fantastical and do not match our intuition include:

The *Heisenberg Uncertainty Principle*: The Heisenberg uncertainty principle states that it is impossible to know the precise location and speed (momentum) of a particle at the same time. It is this indeterminism within nature, especially as interpreted using the Copenhagen interpretation, which Einstein at first refused to accept when he famously stated that "Quantum mechanics is certainly imposing. But an inner voice tells me that it is not yet the real thing. The theory says a lot, but does not really bring us any closer to the secret of the old one. I, at any rate, am convinced that He does not throw dice." [24]

*Quantum Entanglement*: Quantum entanglement occurs when two particles interact in a way that cannot be described independently. Measurement of a property of one of the particles affects the state of the other particle, which at the time of measurement may be separated by a large distance. Again, this quantum mechanical effect goes against our intuition and appears magical, inspiring Einstein to call it "spooky action at a distance." [25]

*Wave Particle Duality*: In quantum mechanics particles can behave both as particles and as waves. As we saw with one-particle interference the way they act seems to depend entirely on if we are watching them or not. Another example where this occurs is in the double slit experiment where particles are fired at a diffraction grating through which they diffract – acting like waves. However, if we observe the particles as they go through the barrier, they behave exactly as we would expect — i.e. like particles. Wave-particle duality was described by Feynman as the quantum "mystery which cannot go away." [26]

## Conclusion

As we have seen, naïve theories of physics can result in incorrect notions about how the world works; and everyone, including physicists, possess these biases that can constrain true objectivity when evaluating scientific explanations [27]. These biases can result in some scientific theories suffering from *bias neglect*, whereby incorrect interpretations of scientific results appear to be objective conclusions. When operating unconsciously under biases, both learned and innate, scientists often draw satisfying conclusions but "robustly fail to appreciate that they should also be more skeptical of such [gratifying] results." [28]

We have illustrated how quantum mechanics, and more specifically the Copenhagen interpretation of quantum mechanics, violates our innate understanding of how the world should work, see table 1. It is our argument, that to satisfy our innate beliefs, many physicists have chosen to support the minimally counter-intuitive and cognitively optimal [29, 30] many-worlds interpretation. In doing so, they do not have to deal with wave-function collapse or the underlying counter-intuitive non-locality that the Copenhagen interpretation espouses.

| Core Object Knowledge | Quantum Mechanics |
| --- | --- |
| Objects do not go in and out of existence. | Particles do go in and out of existence. |
| Objects do not break up and then come back together again. | Particles do break up and then come back together again. |
| Objects cannot teleport from one part of the room to another part without being seen crossing in between. | Particles can teleport from one part of the room to another part without being seen crossing in between. |
| Objects only move when something else makes them move. | Particles are in constant motion |

Table 1. Spelke's Core Object Knowledge and Quantum Mechanics

When determining how to interpret quantum mechanics, physicists should be cognizant of inherent cognitive biases. Otherwise they may adhere to the minimally counter-intuitive many-worlds interpretation as a results of such biases. In fact, we posit that some physicists are biased toward the many-world interpretation exactly because of their cognitive bias arising from their core object knowledge in their naïve physics cognitive module. This partiality does not necessarily mean that these physicists are incorrect; they are simply biased.

**CONFLICT OF INTEREST**
The authors confirm that this article content has no conflicts of interest.

**ACKNOWLEDGMENTS**
The authors would like to thank Bruce Hood whose book [17] and conversations at SciFoo sparked the ideas found in this paper.